\documentclass[12pt,preprint]{aastex}
\usepackage{amssymb,amsmath,bm,graphicx,color}
\newcommand{\maxi}{{\itshape MAXI}}
\newcommand{\asca}{{\itshape ASCA}}

\newcommand{\suzaku}{{\itshape Suzaku}}
\newcommand{\rate}{counts cm$^{-2}$ s$^{-1}$ }

\newcommand{\flux}{ergs cm$^{-2}$ s$^{-1}$}

\newcommand{\colden}{cm$^{-2}$}
\newcommand{\src}{Mrk 421}

\slugcomment{}
\shorttitle{\maxi\ investigation into the longterm X-ray variability of \src}
\shortauthors{N. Isobe et al.}

\begin{document} 
\title{\maxi\ investigation into the longterm X-ray variability 
from the very-high-energy $\gamma$-ray blazar \src}

\author{ 
Naoki      Isobe      \altaffilmark{1,2}, 
Ryosuke    Sato       \altaffilmark{2}, 
Yoshihiro  Ueda       \altaffilmark{2},
Masaaki    Hayashida  \altaffilmark{2},
Megumi     Shidatsu   \altaffilmark{2},
Taiki      Kawamuro   \altaffilmark{2},
Shiro      Ueno       \altaffilmark{3},
Mutsumi    Sugizaki   \altaffilmark{4},
Juri       Sugimoto   \altaffilmark{4},
Tatehiro   Mihara     \altaffilmark{4},
Masaru     Matsuoka   \altaffilmark{3,4},
Hitoshi    Negoro     \altaffilmark{5},
and the \maxi\ team}
\altaffiltext{1}{Institute of Space and Astronautical Science (ISAS), \\
        Japan Aerospace Exploration Agency (JAXA) \\ 
        3-1-1 Yoshinodai, Chuo-ku, Sagamihara, Kanagawa 252-5210, Japan}
\email{n-isobe@ir.isas.jaxa.jp}
\altaffiltext{2}
        {Department of Astronomy, Kyoto University, 
        Kitashirakawa-Oiwake-cho, Sakyo-ku, Kyoto 606-8502, Japan}
\altaffiltext{3}
        {ISS Science Project Office, 
         Institute of Space and Astronautical Science (ISAS), 
         Japan Aerospace Exploration Agency (JAXA), 2-1-1 Sengen,
         Tsukuba, Ibaraki 305-8505, Japan}
\altaffiltext{4}
        {\maxi\ team, Institute of Physical and Chemical Research (RIKEN), 
         2-1 Hirosawa, Wako, Saitama 351-0198, Japan}
\altaffiltext{5}
        {Department of Physics, Nihon University, 1-8-14 Kanda-Surugadai,
         Chiyoda-ku, Tokyo 101-8308, Japan}


\begin{abstract}
The archetypical very-high-energy $\gamma$-ray blazar \src\ 
was monitored for more than 3 years with the Gas Slit Camera 
onboard Monitor of All Sky X-ray Image (\maxi),
and its longterm X-ray variability was investigated.
The \maxi\ lightcurve in the $3$ -- $10$ keV range was 
transformed to the periodogram
in the frequency range $f = 1 \times 10^{-8}$ -- $2 \times 10^{-6}$ Hz.
The artifacts on the periodogram, 
resulting from data gaps in the observed lightcurve, 
were extensively simulated for variations 
with a power-law like Power Spectrum Density (PSD). 
By comparing the observed and simulated periodograms, 
the PSD index was evaluated as $\alpha = 1.60 \pm 0.25 $. 
This index is smaller than that obtained in the higher frequency range 
($f\gtrsim 1 \times 10^{-5}$ Hz), namely,  
$\alpha = 2.14 \pm 0.06$ in the 1998 \asca\ observation of the object. 
The \maxi\ data impose a lower limit on the PSD break 
at $ f_{\rm b} = 5 \times 10^{-6}$ Hz, 
consistent with the break of $f_{\rm b} = 9.5 \times 10^{-6}$ Hz, 
suggested from the \asca\ data.
The low frequency PSD index of \src\ derived with \maxi\
falls well within the range of the typical value 
among nearby Seyfert galaxies ($\alpha = 1$ -- $2$).  
The physical implications from these results are briefly discussed. 
\end{abstract}

\section{Introduction} 
\label{sec:intro}
Blazars, including BL Lacertae objects, 
are one of the most energetic objects
among various classes of active galactic nuclei.
They exhibit rapid and high-amplitude intensity bursts, 
known as flares and/or outbursts. 
Based on the detection of the superluminal motion 
\citep[][]{Superluminal-1,Superluminal-2},
blazars are believed to host a jet emanated at relativistic velocity
along our line of sight. 
The typical Lorentz factor of the jets ejected from BL Lacertae objects 
is estimated as $\Gamma =  5$ -- $30$ \citep[][]{Lorents_Factor}. 
Due to the relativistic beaming effect,
jet emissions dominate the electromagnetic radiation from the blazars.
Their spectral energy distribution is mainly contributed by 
two pronounced components \citep[e.g.,][]{blazar_sequence}. 
The low-frequency component, which peaks in the infra-red to X-ray band,
exhibits strong radio and optical polarization.
Thus, this component is attributed to synchrotron radiation 
from electrons accelerated within the jet.
The high-frequency component comprises X-ray to $\gamma$-ray photons,
generated by inverse Compton scattering of the accelerated electrons 
colliding with soft seed photons. 
The seed photons might be sourced from the synchrotron photons within the jet 
\citep[the synchrotron-self-Compton model; e.g.,][]{IC_component}, 
infra-red photons from the dusty torus \citep{IC_IR_torus}, 
optical/UV photons from the accretion disk \citep{IC_Disk_photon}, 
reprocessed photons from the broad line region \citep{IC_BLR},
and so forth.

Recent observations have detected many blazars 
in the $\gamma$-ray regime 
\citep[][]{MAGIC_Blazars,2FGL_catalog,HESS_blazars}.  
For blazars detected at the very-high-energy $\gamma$-rays 
exceeding $\sim 100$ GeV
(hereafter called VHE blazars, but classically referred to as TeV blazars), 
the synchrotron component frequently peaks 
in the X-ray band (e.g., $0.5$ -- $10$ keV),
covered by the previous X-ray satellites, 
such as \asca\ \citep{ASCA} and \suzaku\ \citep{Suzaku}.
In the observer's frame, the estimated electron cooling timescale
around the synchrotron peak is as short as $ \lesssim 10^{4}$ s
\citep{Mrk421_soft-lag}.
Correspondingly, the most rapid, extreme intensity variation
is expected in the X-ray band.
Therefore, X-ray variability of the VHE blazars
presents as an important probe 
of the jet dynamics and associated acceleration/cooling processes.

Continuous $1$-week monitoring by \asca\ for three classical VHE blazars,
\src\ (the redshift $z = 0.031$), Mrk 501 ($z = 0.034$), 
and PKS 2155$-$304 ($z = 0.117$),
provided new insights into the X-ray variability of VHE blazars
\citep{Mrk421_ASCA_1998,Blazar_PSD_ASCA,HBL_sim_SF}. 
The \asca\ observations revealed that 
the structure function \citep[SF;][]{SF} of the X-ray lightcurves 
of these objects commonly breaks over timescales of $\tau \sim 1$ day,
corresponding to a break of the power spectrum density (PSD)
at frequency $f_{\rm b} = 1/\tau \sim 10^{-5}$ Hz.
Above $f_{\rm b}$, the PSDs of the VHE blazars follow 
a power-law (PL) distribution ($\propto  f^{-\alpha}$) with $\alpha = 2$ -- $3$,
steeper than the slope of the typical PSDs from nearby Seyfert galaxies  
\citep[$\alpha = 1$ -- $2$;][]{AGN_variability,Sy_PSD}. 
Assuming a simple internal shock model,
\citet{HBL_InternalShock} ascribed the SF and PSD break 
of the VHE blazars to the shock-crossing timescale within the jet blobs.
They also attributed the steep PSD index at frequencies above the break
to suppressed variation at timescales shorter than
the shock-crossing timescale. 
However, extensive simulations conducted by \citet{Caveats-to-ST} revealed
that the observed break in the SF of the VHE blazars is 
possible to be an artifact introduced by the limited observation time 
(i.e., $\sim 1$ week). 
This result highlights the need for longterm X-ray monitoring of VHE blazars.

Monitor of All Sky X-ray Image \citep[\maxi;][]{MAXI} is 
the first astronomical observatory 
mounted on the Japanese Experiment Module ``Kibo'' 
of the International Space Station (ISS).
Since its activation in the summer of 2009, 
the Gas Slit Camera \citep[GSC;][]{GSC_introduction},
one of two X-ray instruments aboard \maxi,
has detected more than 500 X-ray sources,
over a period exceeding 3 years, including
100 Seyferts and 15 blazars \citep{2nd_MAXI_catalog}.
On account of its high sky coverage and sensitivity in daily observations
\citep[typically $95$\% and 15 m Crab, respectively;][]{GSC_performance}, 
the GSC presents as an ideal tool for investigating longterm blazar activity.

In the present paper, we analyze 
the longterm X-ray lightcurve of the representative VHE blazar \src,
yielded by the 3-year \maxi\ GSC observations.
The multi-wavelength spectral energy distribution of the source 
is well-explained by the synchrotron-self-Compton process 
\citep[][]{Mrk421_SSC}, 
and the synchrotron peak is known to lie immediately below the X-ray band. 
This VHE blazar is one of the brightest extragalactic objects
listed in the second \maxi\ catalog.
Its 3-year-averaged X-ray flux in the 4 -- 10 keV range is 
$1.8 \times 10^{-10}$ \flux\ 
\citep[corresponding to 15 mCrab;][]{2nd_MAXI_catalog}. 
The blazar exhibited several strong X-ray flares 
during the 3-year observation, three of which were quickly alerted 
by \maxi\ \citep{Mrk421_MAXI,Mrk421_flare_MAXI_3}. 
From the \maxi\ lightcurve, the PSD of \src\ was derived 
in the frequency range $f = 10^{-8}$ -- $2 \times 10^{-6}$ Hz, 
and this complements the higher-frequency \asca\ data 
\citep[$f \gtrsim 10^{-6}$ Hz;][]{Mrk421_ASCA_1998,Blazar_PSD_ASCA}.

The remainder of the paper is organized as follows.
The method for deriving the longterm \maxi\ lightcurve 
of \src\ is described in \S \ref{sec:lcurve}.
The periodogram calculated from the lightcurve 
is presented in \S \ref{sec:PSD_analysis}.
In \S \ref{sec:sim_data_gap},
the effects of the data gaps in the observed lightcurve
on the periodogram are investigated through simulations.
Finally, the PSD shape of \src\ is evaluated 
by comparing the observed and simulated periodograms,
and the implications of the resulting PSD are discussed
in \S \ref{sec:discussion}.

\section{\maxi\ lightcurve} 
\label{sec:lcurve}
\subsection{Data screening} 
\label{sec:screening}
The X-ray lightcurve of \src\ was extracted from \maxi\ GSC data
collected between 2009 September 23 and 2012 October 15. 
During this period, the GSC scanned the target
with its operating proportional counters,
for a total exposure of 1.3 Ms. 
The analyzed GSC event files were provided by the \maxi\ team. 
The following criteria \citep{2nd_MAXI_catalog} were imposed 
for data screening. 
Durations with a high non-X-ray background (NXB) level
were rejected by selecting data with ISS latitudes
between $-40^\circ$ and $40^\circ$. 
Events detected near the edge of each proportional counter, 
at a photon incident angle of $\vert\phi\vert >  38 ^\circ$ 
\citep[defined in][]{GSC_introduction}, were also excluded. 
To avoid flux uncertainty introduced by shadowing, 
we masked sky regions within $5^\circ$ from the solar paddles of the ISS, 
and discarded data from these regions.
We filtered out data taken just after the reboot of \maxi, 
when the counter response was reported to be unstable. 
High background periods related to solar flares were eliminated　
by searching for sudden increases in the count rate. 
After filtering, a net exposure of 0.61 Ms was obtained on the source.

The highest accessible PSD frequency 
of the source intensity variation (i.e., the Nyquist frequency)
was determined from  the time resolution of the lightcurve.
We first binned the \maxi\ lightcurve of \src\ 
into $1$, $3$, $7$, $15$, and $30$ days. 
The fitting procedure for measuring the source flux (described below)
failed for a significant fraction of the 1-day time bins,
especially in fainter phases of the source. 
Therefore, 
the 3-day averaged lightcurve was considered
most suitable for the variability study of this source.

The daily total exposure of \maxi\ on a given celestial target 
is known to be highly variable
and depends on the orbital condition of the ISS. 
\src\ is located in a sky region over which the area 
unobservable by the GSC around the orbital pole of the ISS 
intermittently transits with a typical cycle of $\sim70$ days 
\citep[][]{Mrk421_MAXI}. 
Thus, the variability in exposure time 
is especially problematic for this object.
Days of short exposure, 
for which photon statistics were too poor to place meaningful constraint 
on the source flux, were investigated as follows. 
The $3^\circ \times 3^\circ$ image region centered on \src\
was divided into nine $1^\circ \times 1^\circ$ squares, 
and daily events detected in the $3$ -- $10$ keV band were counted. 
A day was flagged as "bad" (i.e., insufficient exposure time), 
if more than one square contained $5$ events or less, 
or if all the $9$ squares contained only $10$ or fewer events.
In constructing the 3-day averaged lightcurve,
we rejected time bins containing $2$ or $3$ bad days. 
If a bin contained only $1$ or $0$ bad days,
all the events detected in the 3 days were 
included to maximize the photon statistics. 
Consequently, the X-ray lightcurve of \src\
was derived from the \maxi\ data in the final screened exposure of 0.56 Ms.

\subsection{X-ray photometry} 
\label{sec:photometry}
The X-ray flux from \src\ was evaluated in the $3$ -- $10$ keV range, 
where the GSC response is well calibrated 
and the NXB level is relatively low. 
The method used to construct the \maxi\ catalog 
\citep{1st_MAXI_catalog,2nd_MAXI_catalog} was adopted for this analysis. 
In brief, the whole sky was divided into 768 $14^\circ \times 14^\circ$ squares 
using the HEALPix package \citep{HEALPix}.
To estimate the X-ray flux of source candidates, 
observed images of individual regions were fitted 
to the corresponding point spread functions (PSFs) plus the NXB model, 
based on the maximum likelihood method with C-statistics \citep{C-stat}. 
The fit was performed on the central $11^\circ \times 11^\circ$ square
to avoid confusion from sources outside the region.

From these 768 squares, 
we selected the area of which the center is closest to \src.  
For each $3$-day time bin, the GSC image of this region 
was accumulated with a spatial resolution of $0.7^\circ$. 
The NXB model images 
were estimated by the \maxi\ simulator \citep{MAXIsim},
accounting for the temporal variability in the NXB level and 
its spatial distribution over the detection counters. 
When generating the PSF at the \src\ location (Table \ref{table:src}), 
a Crab-like X-ray spectrum, with a photon index of $\Gamma = 2.1$ 
and an absorption column of $N_{\rm H} = 0.35 \times 10^{22}$ \colden\ 
\citep{Crab_spec}, 
was assumed as the input source spectrum to the \maxi\ simulator. 
Two possible contaminating X-ray sources 
(Src 1 and Src 2 in Table \ref{table:src}) 
were also found within the \src\ region.
Both sources were omitted from the second \maxi\ catalog,
since their detection significance was below the threshold 
adopted for the catalog \citep[$\sigma = 7$;][]{2nd_MAXI_catalog}; 
however, they were included in the image fitting
for measuring the X-ray flux from \src. 
Both sources are separated from the target by $\sim 5^\circ$.
Since this angular separation exceeds the GSC PSF, 
of which the full width at half maximum is $\sim 1.5^\circ$ 
in the $2$ -- $10$ keV \citep{GSC_performance},
neither source should introduce significant systematic error. 
Adopting these procedures, 
we derived the X-ray lightcurve of \src\ from the $3$-year GSC data 
with a time resolution of 3 days.
The lightcurve is shown in Figure \ref{fig:lightcurve}.
The average and standard deviation of the source count rate, 
weighted by the statistical errors,
were evaluated as $2.0 \times 10^{-2}$ \rate\ and  
$2.6 \times 10^{-2}$ \rate  respectively,
while the unweighted values were derived as 
$2.9 \times 10^{-2}$ \rate\ and $ 3.2 \times 10^{-2}$ \rate.   

As evident in Figure \ref{fig:lightcurve}, 
\src\ widely varied in the $3$ -- $10$ keV range during the \maxi\ monitoring. 
In particular, several prominent X-ray flares were detected from the source.
We have issued a quick alert for three of these flares
(indicated by arrows in Figure \ref{fig:lightcurve}),
occurring on 2010 January 1, February 16, and 2011 September $7$ -- $8$ 
\citep{Mrk421_MAXI,Mrk421_flare_MAXI_3}. 
The lightcurve displays 21 data gaps, 
during which no X-ray flux was measured because 
the time bins failed the screening criteria described 
in \S \ref{sec:screening}.
The duration of the data gap is distributed as shown 
in Figure \ref{fig:gap_hist}. 
The average duration of the gaps is $21.1$ days.

\section{PSD analysis} 
\label{sec:PSD_analysis}
The PSD is commonly used to quantify the amplitude and timescale of 
intensity variation from various classes of astrophysical objects,
including Galactic black hole binaries, 
Seyferts \citep[e.g.,][]{AGN_variability} 
and blazars \citep[e.g.,][]{Blazar_PSD_ASCA}. 
To estimate the PSD of \src,
we adopted the periodogram normalized by the averaged source intensity 
after \citet[][]{GX339} and \citet[][]{AGN_variability};
\begin{eqnarray}
P(f) &=& \frac{ [a^2(f) + b^2(f) ] T }{\mu^2}, \nonumber\\  
a(f) &=& \frac{1}{n} \sum_{j=0}^{n-1} x_j \cos (2\pi f t_j), \nonumber \\
b(f) &=& \frac{1}{n} \sum_{j=0}^{n-1} x_j \sin (2\pi f t_j),
\label{eq:PSD}
\end{eqnarray}
where $f$ is the frequency of the intensity variation,
$x_j$ is the source count rate at time $t_j$ ($ 0 \le j \le n-1 $),
$n$ is the number of data points, 
$T$ ($= t_{n-1} - t_0$) is the total duration of the lightcurve,  
and $\mu$ is the unweighted average of the source count rate.
Based on this definition, 
integration of the periodogram over the positive frequencies 
yields half of the excess variance.
The underlying PSD, i.e., $S(f)$, of the source is evaluated
by fitting a continuous function (such as a PL) to the observed periodogram.
The Poisson level corresponding to the power due to the statistical fluctuation
is calculated as 
\begin{eqnarray}
P_{\rm stat} 
   &=& \frac{T \sigma_{\rm stat}^{2}}{n \mu^2} \nonumber\\ 
\sigma_{\rm stat}^2 
   &=& \sum_{j=0}^{n-1} \frac{(\Delta x_{\rm j})^2 }{n} 
\end{eqnarray}
where $\Delta x_{\rm j}$ denotes the error of the source count rate at $t_j$.

From the \maxi\ lightcurve of \src\ in Figure \ref{fig:lightcurve}, 
we can investigate the X-ray variability in the frequency range 
of $f = 1 \times 10^{-8}$ -- $2 \times 10^{-6}$ Hz,
which has yet to be explored for blazars. 
However, to calculate the periodogram in equation  (\ref{eq:PSD}),
a continuous data stream with regular sampling 
(i.e., no data gaps) is required. 
To achieve this, we interpolated the missing time bin data 
by averaging the data in the five time bins (corresponding to 15 days) 
preceding and succeeding the missing bin (averaging over 10 bins in total).
From these 10 time bins, those overlapping the gaps were discarded. 
The resulting continuous \maxi\ lightcurve of \src\
is plotted as the dashed line in Figure \ref{fig:lightcurve}.
The interpolation procedure exerted only a minor effect on 
the weighted mean and standard deviation, which were re-evaluated 
as $2.0 \times 10^{-2}$ \rate\ and $ 2.3 \times 10^{-2}$ \rate\ respectively,
in addition to the unweighted ones, 
$2.6 \times 10^{-2} $ \rate\ and $ 2.7 \times 10^{-2} $ \rate.

Before investigating the variability of \src,  
we analyzed the \maxi\ lightcurves of two bright galaxy clusters,
the Coma Cluster and Centaurus Cluster,
with X-ray fluxes of $2.5 \times 10^{-10}$ \flux\ and 
$9.3 \times 10^{-11}$ \flux, respectively in the $4$ - $10$ keV range 
listed in the \maxi\ catalog \citep[][]{2nd_MAXI_catalog}. 
The flux of these sources is regarded as constant over the \maxi\ monitoring. 
We constructed the lightcurves and corresponding periodograms of both objects 
as described above.
The periodogram of these sources became consistent 
with the Poisson level (i,e, $P(f) \sim P_{\rm stat}$),
regardless of the variation frequency $f$. 
Therefore, we concluded that our interpolation technique exerted  
no significant systematic impact on the results of stable sources. 

The solid line in the left panel of Figure \ref{fig:psd} 
presents the periodogram of \src\ 
in the frequency range of $f = 1 \times 10^{-8}$ -- $2\times10^{-6}$ Hz
evaluated by equation (\ref{eq:PSD}) from the interpolated \maxi\ lightcurve.
Next, this raw periodogram was binned by a factor of 1.6 in the frequency.
The frequency intervals were additionally bundled, 
such that each bin included at least two data points.
Adopting the procedure proposed by \citet{red_noise},
the average value of the logarithm of the raw periodogram, 
$\overline{\log(P(f))}$ 
(hereafter referred to as the binned logarithmic periodogram), 
was calculated for each bin,
while its representative frequency was evaluated 
by the geometric mean frequency.
Here, the Poisson level ($P_{\rm stat} = 4.56 \times 10^{3}$ Hz$^{-1}$)
was not subtracted,
because the several high-frequency data points are located below it,
as shown in the left panel of Figure \ref{fig:psd}.
The effect of the Poisson level is investigated through the simulation 
(see \S \ref{sec:discussion}),
when the best-fit PSD parameters are determined.
Since the logarithm of the raw periodogram theoretically 
exhibits a constant variance of $\pi^2/6(\ln(10))^2=0.310$ 
(corresponding to the variance of $\log(\chi^2_2/2)$ 
where $\chi^2_2$ is the $\chi^2$ distribution with two degrees of freedom),
the variance of the binned logarithmic periodogram was set 
at $\sigma_{\log(P(f))}^2=0.310/m$,
with $m$ being the number of the data points in the individual bins. 
It is known that the binned logarithmic periodogram is biased as 
$\overline{\log(P(f))} = \log(S(f)) - 0.25068$ 
\citep[e.g.,][]{Vaughan},
where the constant offset of $-0.25068$ 
equals to the average of the $\log(\chi^2_2/2)$ distribution.
Throughout the present paper, 
this offset is corrected by adding 0.25068 to $\overline{\log(P(f))}$.
The right panel of Figure \ref{fig:psd} displays 
the binned logarithmic periodogram.
Over this frequency range, the periodogram appears to approximate a PL form.

\section{PSD artifacts introduced by data gaps} 
\label{sec:sim_data_gap}
The data gaps in the observed lightcurve can interfere with our PSD estimate. 
As shown in Figure \ref{fig:gap_hist}, they persist over $3$ -- $39$ days,
corresponding to frequencies of $f = (0.3 - 4) \times 10^{-6}$ Hz.
Above these frequencies the variability power may be over/underestimated 
by interpolating the data gaps.
We thus investigated the influence of the gaps, 
using the simulation approach.

As described in \citet[][TK95]{PSD_Simulation}, 
we produced 1000 simulated lightcurves in 3-day bins 
with PL-type underlying PSDs of the form $ S(f) \propto f^{-\alpha_0}$.
We, here, assumed an ideal case without any Poisson fluctuation
(corresponding to $P_{\rm stat} = 0$).  
A sampling pattern that exactly mimicked the observation was applied 
to the simulated data sets, yielding discontinuous artificial lightcurves.
Following the processing procedure of the observed \maxi\ data, 
the fluxes in the gaps were interpolated 
and a periodogram, $P_{\rm s}(f)$, was derived from each simulated lightcurve. 
In the identical manner to that for the observed data,
the binned logarithmic periodogram, $\overline{\log(P_{\rm s}(f))}$,
was computed for each simulated periodogram.
From the ensemble of the 1000 simulated products,
the average and standard deviation of the binned logarithmic periodogram,
$\langle \overline{\log{P_{\rm s}(f)}} \rangle$ and 
$\sigma_{ \overline{ \log{P_{\rm s} (f) } } }$ respectively,
were evaluated for the individual frequency bins.
By directly comparing $\overline{\log(P(f))}$ and 
$\langle \overline{\log{P_{\rm s}(f)}} \rangle$,
we canceled out the effect of the data gaps. 
To correct for the effects of the red-noise leak \citep[][]{red_noise}, 
the transfer of power from lower to higher frequencies 
introduced by the limited observation period,
lightcurves of 10-times-longer duration were initially created
and divided into segments of 10 data sets. 
These processes were repeated for different values of the PL index 
$\alpha_0$.

\begin{figure*}[h]
\plottwo{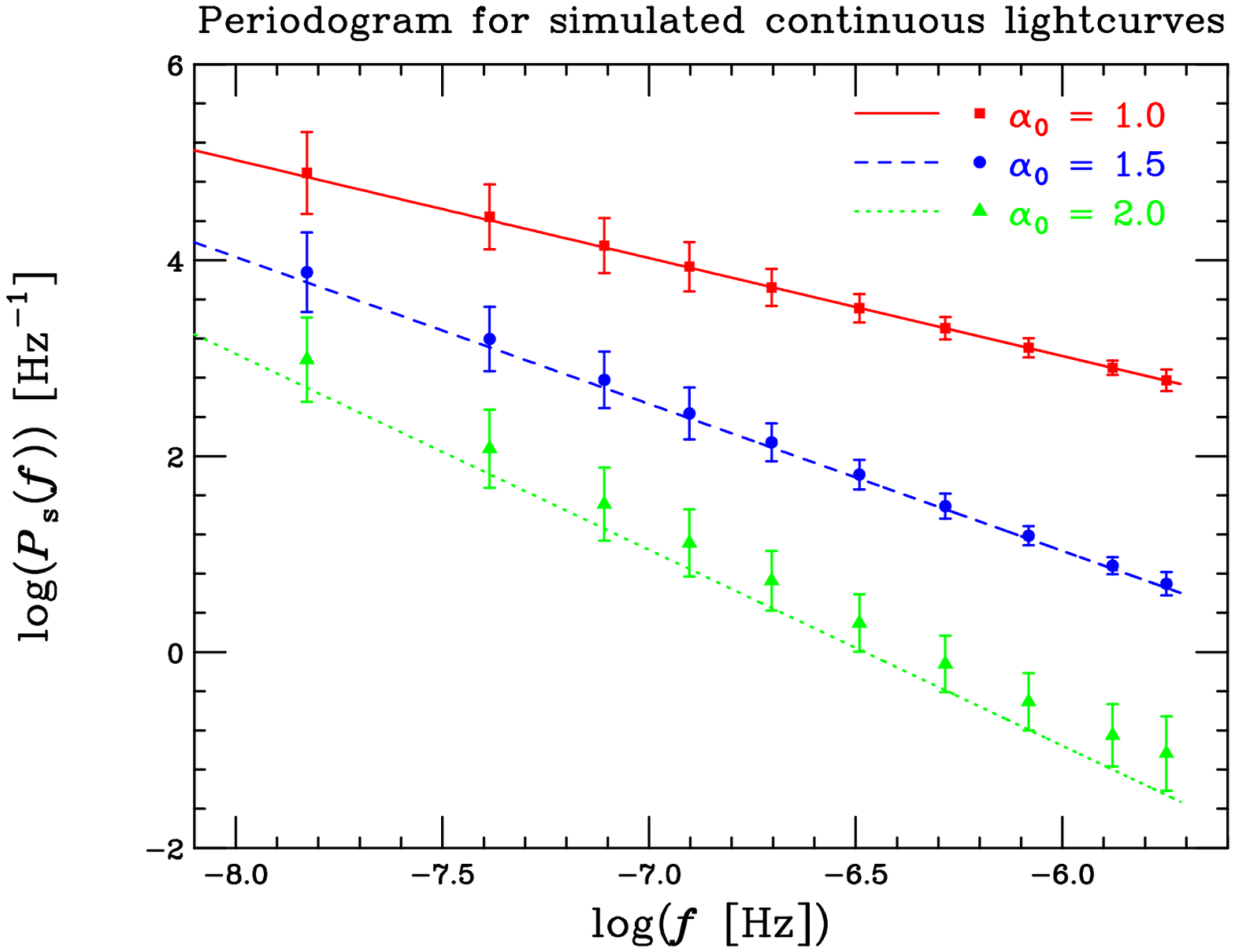}{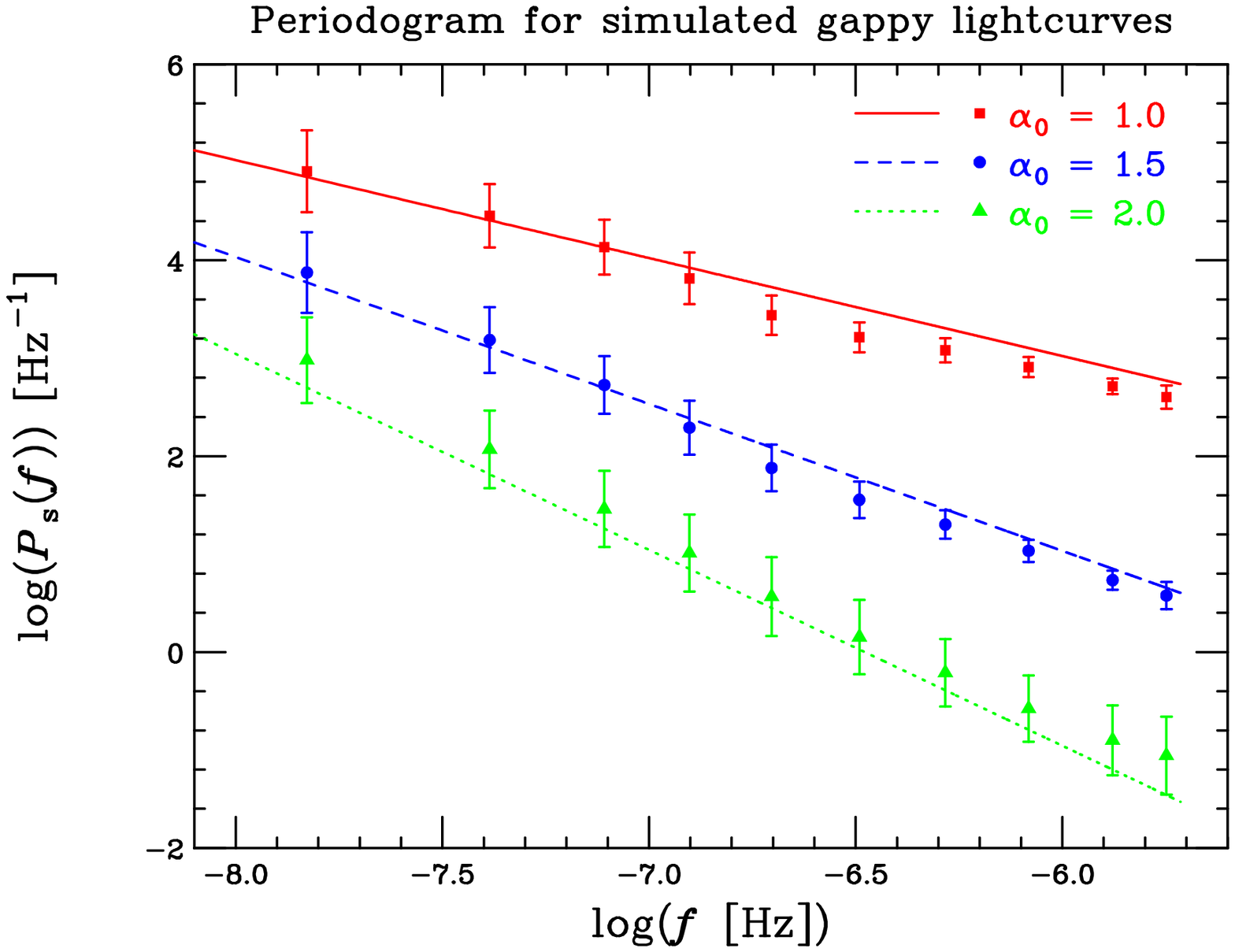}
\caption{
(Left) Binned logarithmic periodograms 
($\langle \overline{\log{P_{\rm s}(f)}} \rangle$ 
and $\sigma_{ \overline{ \log{P_{\rm s} (f) } } }$)
derived from the simulated continuous lightcurves without any data gaps, 
in comparison with the corresponding PSD model, $S(f)$. 
(Right) Binned logarithmic periodograms 
for the simulated discontinuous data sets, accounting for the data gaps.
The results for the input PSD index $\alpha_0 = 1.0$, $1.5$ and $2.0$
are indicated by the red squares, blue circles, and green triangles, 
respectively, while the PSD models 
are plotted as the solid, dashed and dotted lines, respectively.
}
\label{fig:sim_psd}
\end{figure*}

The left panel of Figure \ref{fig:sim_psd} compares 
$\langle \overline{\log{P_{\rm s}(f)}} \rangle$ 
derived from the simulated gapless data set
with the PSD input to the simulations, $S(f) \propto f^{-\alpha_0}$,
for the PSD index of $\alpha_0 = 1.0$, $1.5$ and $2.0$. 
The binned logarithmic periodogram from the simulation is found 
to coincide with the underlying PSD model for $\alpha_0 = 1.0$ and $1.5$,
but overestimates the input PSD for $\alpha_0 = 2.0$, 
especially in the range of $f \gtrsim 5 \times 10^{-7}$ Hz.
The offset of $\langle \overline{\log{P_{\rm s}(f)}} \rangle$ from $\log (S(f))$
is attributable to red-noise leak,
whose impact is predicted to be more significant 
at higher PSD indices \citep{red_noise}.

The right panel of Figure \ref{fig:sim_psd} plots 
$\langle \overline{ \log{P_{\rm s}(f)} } \rangle$
derived from the simulations containing the data gaps. 
For the PSD indices $\alpha_0 = 1.0$ and $1.5$,
for which the red-noise leak has a minor impact,
the simulated periodogram clearly underestimates the input PSD $S(f)$,
over the frequency range $f \sim 1 \times 10^{-7}$ -- $1 \times 10^{-6}$ Hz. 
This frequency range corresponds to the timescale of the data gaps 
($3$ -- $39$ days; Figure \ref{fig:gap_hist}).
After the binned logarithmic periodogram was converted 
to the linear value as $10^{\langle \overline{ \log{P_{\rm s}(f)} } \rangle}$,
the ratio $R_1$ was evaluated 
between the simulated products with and without the data gaps.
The left panel of Figure \ref{fig:psd_ratio} indicates that
the data gaps appear to artificially diminish 
the variability power around $f = (2$--$4) \times 10^{-7}$ Hz
by a factor of up to $R_1 \sim 1/2$. 
The right panel of Figure \ref{fig:psd_ratio} displays 
the ratio $R_2$ of the periodogram derived from 
the simulated discontinuous lightcurves to the underlying PSD,
defined as $ R_2 = 10^{\langle \overline{ \log{P_{\rm s}(f)} } \rangle}/S(f)$. 
In these results, 
the periodogram is less affected by the data gaps
at frequencies $f \lesssim 1 \times 10^{-7} $ Hz,
where the ratio ranges are 
$0.7 \lesssim R_1 \lesssim 1.05$ and $0.8 \lesssim R_2 \lesssim 1.2$
for $\alpha \lesssim 1.5 $.

\section{Discussion} 
\label{sec:discussion}
\subsection{Comparison between the \maxi\ and \asca\ data} 
\label{sec:comp_maxi_asca}
Figure \ref{fig:comp_maxi-asca} plots the periodogram of \src\,
multiplied by the frequency (i.e., in the $f P(f)$ form)
in the frequency range $f = 1 \times 10^{-8}$ -- $ 2\times10^{-6}$ Hz,
derived from the 3 -- 10 keV \maxi\ lightcurve.
The \maxi\ periodogram is uncorrected for artifacts 
arising from data gaps and red-noise leak 
(investigated in \S \ref{sec:sim_data_gap}).
Here, we formally subtracted the Poisson level, 
after the binned logarithmic periodogram 
was translated into the linear value.
For comparison, the periodogram of the source 
over the higher frequency range of $f = 10^{-6}$ -- $10^{-3}$ Hz,
observed by \asca\ in the $0.7$ -- $7.5$ keV range, is also shown. 
Here, we adopt the \asca\ data collected during the 1998 observation,
of which the duration ($\sim 1$ week) is the longest 
among the three \asca\ observations reported by \citet{Blazar_PSD_ASCA}.
The periodograms derived from the \maxi\ and \asca\ data 
were not modified by any normalization scaling factors. 
It is important to note that 
a factor of 4 change was recorded in the variability power 
among the three \asca\ observations. 

\citet{Mrk421_soft-lag} reported that 
the shape of the SF of the \src\ variability in the $0.07$ -- $7.5$ keV range
is energy independent from $10^{-2}$ to $3$ days.
This suggests that the PSD shape of \src\ is independent of the photon energy
over the frequency range $f = 10^{-6}$ -- $10^{-3}$ Hz.
Accordingly, we assumed that 
the PSD shape in the \maxi\ energy range ($3$ -- $10$ keV) 
is similar to that in the \asca\ range ($0.7$ -- $7.5$ keV).
At $2$ -- $7.5$ keV, the observed variability power of the source 
is half that at $0.5$ -- $2$ keV \citep{Mrk421_soft-lag}. 
Consequently, the PSD normalization is possible to differ
between the \maxi\ and \asca\ energy ranges by a factor of 2, at most.

\citet{Blazar_PSD_ASCA} reported that the periodogram of \src\ 
above $f= 10^{-5}$ Hz, derived from the 1998 \asca\ observation, 
was reproduced  by a PL-like PSD model 
with a slope of $\alpha_{\rm H} = 2.14 \pm 0.06$.
They also revealed a PSD break at $f_{\rm b} = (9.5 \pm 0.1) \times 10^{-6}$ Hz 
(as indicated by the blue dotted line in Figure \ref{fig:comp_maxi-asca}). 
Although the error was large, 
the low frequency PSD index (below $f_{\rm b}$) was estimated 
as $\alpha_{\rm L} = 0.88 \pm 0.43$. 
In the remaining two \asca\ data sets reported by \citet{Blazar_PSD_ASCA},
the short observation duration 
($\sim 1$ day corresponding to $f \gtrsim 10^{-5}$ Hz) 
precluded detection of a PSD break in these observations.

The \maxi\ data are critically important in constraining 
the PSD shape of the source at lower frequencies ($ f < 10^{-5} $ Hz).
As shown in Figure \ref{fig:comp_maxi-asca},  
the periodogram derived from the \maxi\ data 
below $f = 2 \times 10^{-6}$ Hz appears flatter than 
that derived from \asca\ above $f_{\rm b}$.
In addition, both data sets indicate a similar variability power 
around $f \gtrsim 2 \times 10^{-6}$ Hz.
Thus, the \maxi\ data appear to qualitatively support the PSD break
around the boundary between the \maxi\ and \asca\ frequency ranges.
However, to determine the underlying PSD shape in this frequency range,
the artifacts of the data gaps must be deconvolved from the \maxi\ data 
(\S \ref{sec:PSDindex}).

Similar to the analysis in \S \ref{sec:sim_data_gap},
we simulated the \maxi\ periodogram for discontinuous lightcurves, 
by simply extending the best-fit PL PSD model to the \asca\ data 
above the frequency break 
toward the lower frequency range of $f < 2 \times 10^{-6}$ Hz. 
The best-fit PSD index and normalization to the \asca\ data was 
$\alpha_{\rm H} = 2.14$ and $N_{\rm 0} = 8.6 \times 10^{2}$ Hz$^{-1}$ 
at $f_{\rm b} = 9.5 \times 10^{-6}$ Hz, respectively \citep{Blazar_PSD_ASCA}. 
The periodogram simulated by this model 
indicated by filled triangles in Figure \ref{fig:comp_maxi-asca} 
clearly exceed the observed data points over the entire \maxi\ frequency range. 
Even accounting for possible systematic PSD normalization factors 
between the \maxi\ and \asca\ data sets, as discussed above, 
the simulated and actual \maxi\ data significantly differ.
This implies that the PL-like PSD model fitted to the \asca\ data 
above $f > 10^{-5}$ Hz (i.e., $\alpha_{\rm H} = 2.14$) 
does not smoothly extend  into the \maxi\ frequency range.
Thus, we roughly estimated a lower limit on the PSD break frequency 
as $f_{\rm b} \gtrsim 2 \times 10^{-6}$ Hz, 
and successfully reinforced the \asca\ result.

\subsection{Low-frequency PSD Index} 
\label{sec:PSDindex}
Next, we investigated the PSD shape of \src\ 
in the frequency range $f = 10^{-8}$ -- $2 \times 10^{-6}$ Hz
by comparing the \maxi\ data with 
the simulated results 
in which the effects of the gap and red-noise leak were considered.
The Poisson level was also included in the simulated periodograms,
instead of subtracting it from the observed binned logarithmic periodogram.
Over the \maxi\ range, 
the underlying PSD of the object was assumed to follow a simple PL function.
The free parameters to be constrained in this model are 
the PSD normalization $N_{\rm 0}$ and the index $\alpha_0$.
The normalization is evaluated at the putative break frequency 
$f_{\rm b} = 9.5 \times 10^{-6}$ Hz suggested from the \asca\ data. 
Thus, the PSD function input to the simulation was 
$S(f) = N_0 (f/f_{\rm b})^{-\alpha_0}$. 
Following the standard manner, 
the best-fit $N_{\rm 0}$ and $\alpha_0$ values were obtained 
by minimizing the $\chi^2$ function defined as 
\begin{eqnarray}
\label{eq:chi2_mod}
\chi^2 = 
\sum_{j}^{} \left( 
    \frac{\overline{\log(P(f_j))}-\langle\overline{\log{P_{\rm s}(f_j)}}\rangle }
         { \sigma_{\log(P(f_j))} } 
\right)^2
\end{eqnarray}
where $f_j$ denotes the frequency of $j$-th bin 
of the binned logarithmic periodogram
with $j = 0$ and $9$ corresponding to 
the lowest and highest frequency ones, respectively.

\begin{figure*}[h]
\plottwo{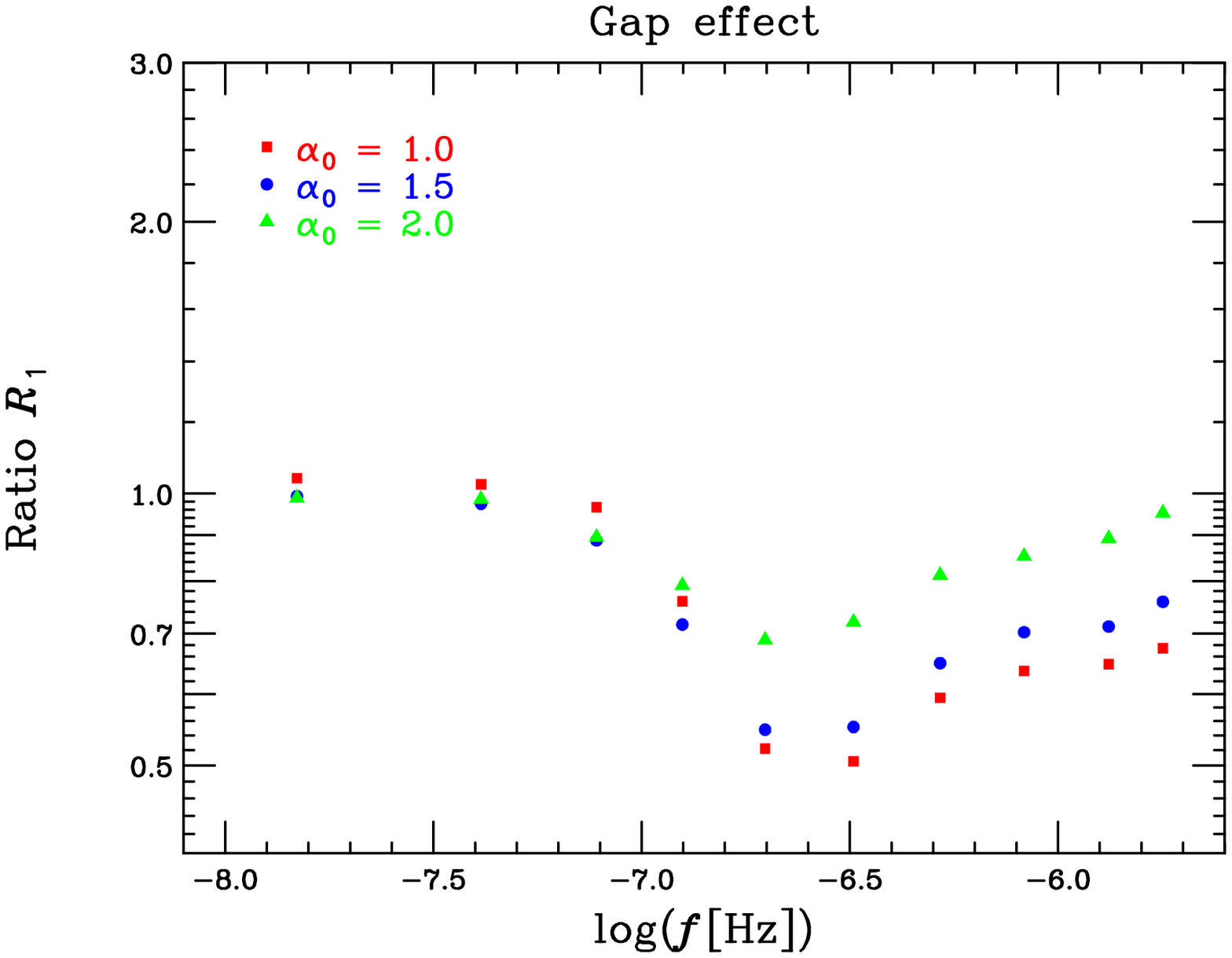}{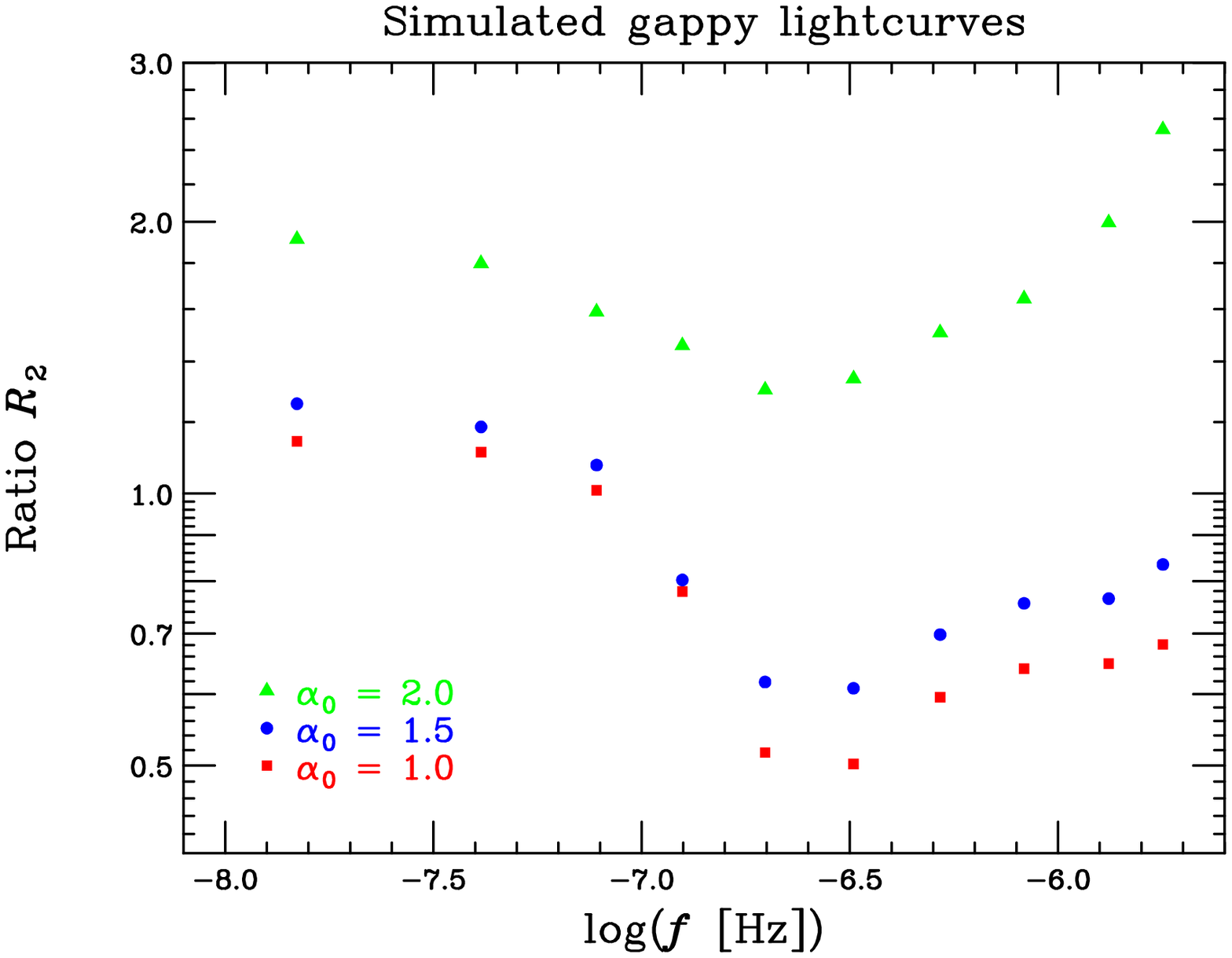}
\caption{
(Left) Ratio $R_1$ between the simulated periodograms 
considering and ignoring data gaps.  
Red squares, blue circles, and green triangles, show the results 
for input PSD indices $\alpha_0 = 1.0$, $1.5$, and $2.0$, respectively.
(Right) Ratio $R_2$ of the periodogram from the simulated discontinuous 
lightcurves to the input PSD.
}
\label{fig:psd_ratio}
\end{figure*}

First, we attempted to constrain $\alpha_0$ and $N_0 $,
using all the 10 data points in the \maxi\ binned logarithmic periodogram,
displayed in Figure \ref{fig:psd}.
However, the fit was unsuccessful within the range 
$\alpha_0 = 0.5$ -- $2.5$ and $N_0 = 10$ -- $10^{5}$ Hz$^{-1}$
($\chi^2/{\rm dof} \ge 23.5 / 8$), 
mainly because the \maxi\ data exhibit a hump feature
over the $(2$ -- $8) \times 10^{-7}$ Hz range.
Using data collected by 
the All-Sky Monitor onboard the Rossi X-ray Timing Explorer,
\citet{Mrk421_PSD_peak} reported a possible peak 
in the PSD of \src\ at $f = 1.9 \times 10^{-7}$ Hz, 
but did not rigorously evaluate its significance. 
Although the hump feature in the \maxi\ data resembles 
that of a quasi-periodic oscillation, 
its frequency exactly coincides with the frequency range of the artifacts
introduced by the data gaps (Figure \ref{fig:psd_ratio}).  
We found that the hump cannot be readily disentangled
from the gap artifacts using the \maxi\ data alone.
Therefore, this feature is not further discussed in the present paper.

\begin{figure}[h]
\plotone{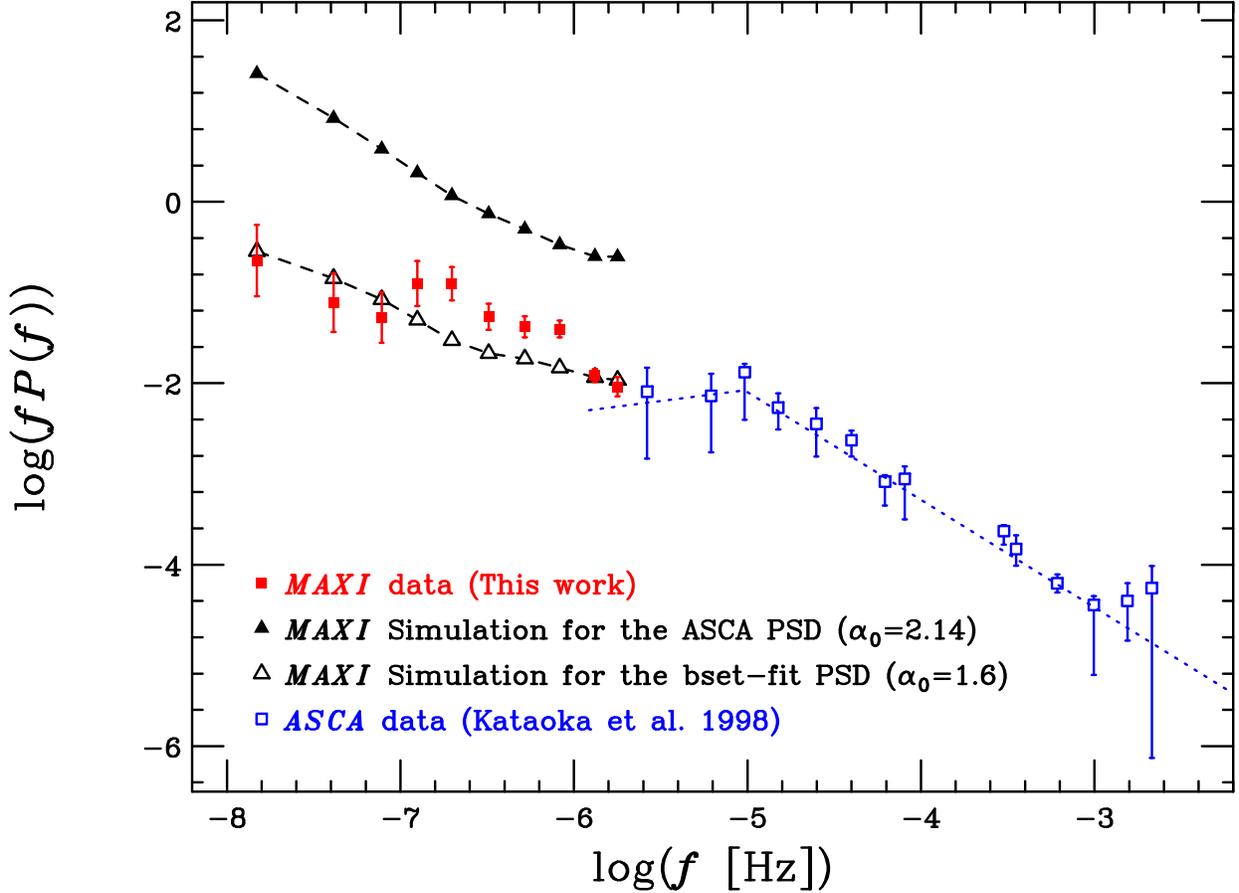}
\caption{
Comparison between the \maxi\ and \asca\ periodograms,
which are multiplied by frequency,
shown in the logarithmic space (i.e., in the $\log( f P(f))$ form).
The red filled squares indicate the \maxi\ data points
(evaluated as $\log(f) + \overline{\log(P(f))}$),
which were uncorrected for the artifacts introduced 
by the data gaps and red-noise leak (see \S \ref{sec:sim_data_gap}). 
The Poisson level was subtracted.
The blue open squares are taken from 
the long-look \asca\ observation in 1998 \citep{Blazar_PSD_ASCA}.
The \asca\ data were reported to be successfully reproduced by
a broken-PL PSD model (dotted line).
The filled triangles show the simulated \maxi\ periodogram 
($\log(f) + \langle \overline{\log{P_{\rm s}(f)}} \rangle$)
when the PL PSD model with $\alpha_0 = 2.14$
that best-fitted the \asca\ data above the break frequency 
($f_{\rm b} = 9.5 \times 10^{-6}$ Hz) 
was simply extrapolated down to $f =  1 \times 10^{-9}$ Hz.
The open triangles indicate the periodogram simulation of 
the best-fit PL model to the 6 \maxi\ data points ($\alpha_0 = 1.60$).
All the simulation results account for the data gaps and red-noise leak.
}
\label{fig:comp_maxi-asca}
\end{figure}

\begin{figure*}[h]
\plotone{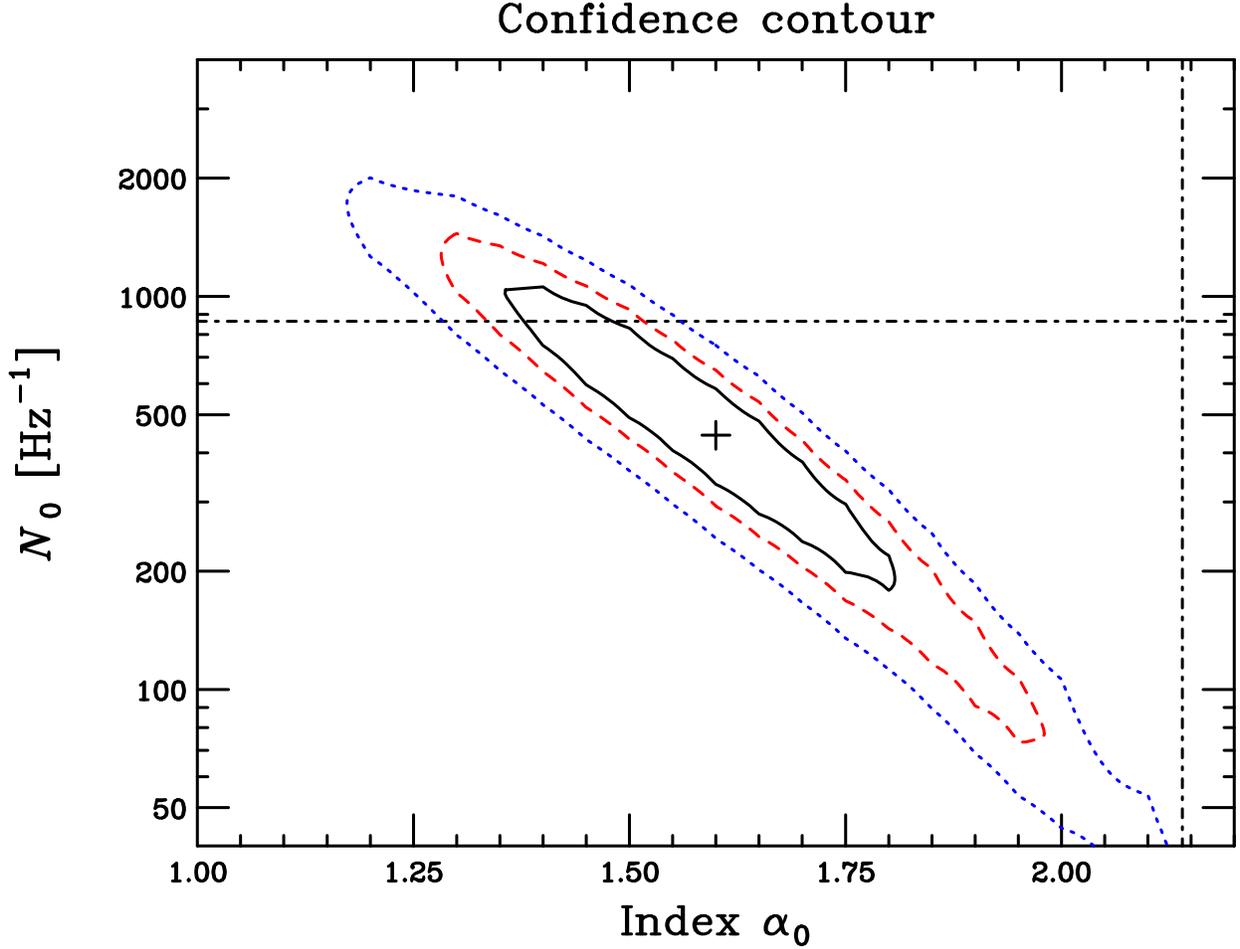}
\caption{
Confidence map on the plane of the PSD normalization $N_{\rm 0}$
and PSD index $\alpha_{\rm 0}$,
derived from the 6 \maxi\ data points. 
The PSD normalization $N_0$ was evaluated at the possible break point
suggested from the \asca\ observation, 
$f_{\rm b} = 9.5 \times 10^{-6}$ Hz \citep{Blazar_PSD_ASCA}. 
The cross indicates the best-fit $N_{\rm 0}$ and $\alpha_{\rm 0}$ values 
($\chi^2/{\rm dof} = 4.0/4$).
Three contours are drawn at the confidence levels of $68$\%, $90$\% and $99$\% 
(i.e., corresponding to $\Delta \chi^2 = 2.30$, $4.61$ and $9.21$)
with the black solid, red dashed and blue dotted lines, respectively. 
The horizontal and vertical dash-dotted lines 
indicate the best-fit values to the \asca\ data above the break 
\citep[$N_0 = 8.6 \times 10^2$ Hz$^{-1}$ and $\alpha_{\rm H} = 2.14$;][]
      {Blazar_PSD_ASCA}. 
}
\label{fig:prob_reject}
\end{figure*}

To reduce the effect of the data-gap artifacts
and to avoid possible contamination from the hump structure,  
we next evaluated the PSD shape from the 6 data points 
with $j= 0$ -- $3$, $8$ and $9$, 
where the periodogram appears to be less affected by the gaps 
(i.e., where $R_1 \gtrsim 0.7$ for $\alpha_0 \sim 1.5$; 
left panel of Figure \ref{fig:psd_ratio}). 
Figure \ref{fig:prob_reject} displays the confidence contours 
on the $N_{\rm 0}$--$\alpha_0$ plane. 
The best-fit to the \maxi\ data was achieved at a PSD index and normalization
$\alpha_0 = 1.60 \pm 0.25 $ and $N_0 = 4.4_{-3.0}^{+7.3} \times 10^2$ Hz$^{-1}$, 
respectively, ($\chi^2/{\rm dof} = 4.0/4$), 
where the errors were evaluated at the 90\% confidence level.
The binned logarithmic periodogram
simulated for the best-fit PSD model are overlaid 
on the right panel of Figure \ref{fig:psd} with the open triangles.
It is also plotted in Figure \ref{fig:comp_maxi-asca},
after the Poisson level was subtracted 
in the similar manner to the observed data.

The variability power at the possible break determined from \asca,
$N_{\rm 0} = 8.6 \times 10^2$ Hz$^{-1}$ \citep{Blazar_PSD_ASCA}, 
is within the acceptable region of the \maxi\ data. 
Fixing the normalization at the \asca\ value, 
the PSD index in the \maxi\ range 
is estimated as $\alpha_0 = 1.45_{-0.15}^{+0.10}$. 
Accounting for the possible uncertainty in $N_0$ 
discussed in \S \ref{sec:comp_maxi_asca} 
(a factor of 4 caused by PSD fluctuations in the \asca\ observations 
and a factor of at most 2 introduced by the different energy coverage), 
the maximum systematic error in $\alpha_0$ is $\Delta \alpha_0 \sim 0.4$. 
Therefore, we infer that the PSD slope in the \maxi\ frequency range 
is flatter than that above the break measured from the \asca\ data
($\alpha_{\rm H} = 2.14 \pm 0.06 $),
but is consistent with the \asca\ PSD slope below the break 
($\alpha_{\rm L} = 0.88 \pm 0.43 $), albeit with rather large errors. 
In the following discussion, 
we adopt the \maxi\ result for the low frequency PSD index of \src,
since the \maxi\ data cover a wider frequency range below the break. 

These results provide a quantitative support for 
the PSD break between the \maxi\ and \asca\ frequency range. 
Surveying the $N_{\rm 0}$--$\alpha_0$ plane,
we found that the PL model acceptable to the \maxi\ data 
intersects the extrapolation of the \asca\ result 
above $f > 1 \times 10^{-5}$ Hz in the $f \ge 5 \times 10^{-6}$ Hz  range.
This implies that the break suggested from the \maxi\ data 
is absolutely consistent with the ASCA result \citep{Blazar_PSD_ASCA}. 
Therefore, we concluded that the PSD of the X-ray variability of \src\ 
breaks at $ f_{\rm b} = (5$ -- $ 9.5) \times 10^{-6}$ Hz. 
These results clearly demonstrate the utility of \maxi\ data 
for evaluating the longterm variability of blazars and,
presumably, that of other classes of active galactic nuclei. 

Investigating the X-ray lightcurves of the three archetypical VHE blazars
(including \src), 
derived from \asca\ long-look observations (exceeding $\sim 1$ week),
\citet{HBL_sim_SF} identified a break in the SF 
at timescales of $\tau \sim 1$ day in all three blazars. 
In this case, the SF break at $\tau \sim 1$ day corresponds 
to a PSD break of $f_{\rm b} = 1/ \tau \sim 1 \times 10^{-5}$ Hz,
almost consistent with our result. 
However, \citet{Caveats-to-ST}, who conducted detailed lightcurve simulations,
proposed that the SF break reported in \citet{HBL_sim_SF} 
was artificially induced by the \asca\ observation period 
(i.e., $\gtrsim 1$ week). 
In contrast to the SF, 
the periodogram is reported to be relatively robust to artificial breaks
introduced by truncated data lengths \citep{Caveats-to-ST}.
In \S \ref{sec:sim_data_gap}, we also carefully simulated 
the artifacts from the data sampling window introduced 
to the \maxi\ periodogram.
These analyses confirmed that the PSD of \src\ breaks 
at $ f_{\rm b} = (5$ -- $ 9.5) \times 10^{-6}$ Hz
(corresponding to the SF break at $\tau \sim $ a few days), 
where the results were derived by combining the \maxi\ and \asca\ data sets.

Recently, \citet{Mrk421_BAT_PSD} investigated 
the hard X-ray lightcurve of \src\ in the $14$ -- $150$ keV energy range,
extracted from 58-month observations 
by the {\itshape Swift} Burst Alert Telescope (BAT)\footnote{{\tt
http://heasarc.nasa.gov/docs/swift/results/bs58mon/SWIFT\_J1104.4p3812}}.
They derived a hard X-ray PSD index of $\alpha = 0.85 \pm 0.25$
in the frequency range $f = 10^{-8}$ -- $10^{-6}$ Hz,
significantly smaller than the \maxi\ result 
of $\alpha_0 = 1.60 \pm 0.25$ in $3$ -- $10$ keV.
This result indicates that the X-ray variability pattern  
of \src\ is energy-dependent especially toward the harder X-ray band,
although no energy dependence was suggested 
in the PSD shape below $\sim 10$ keV \citep{Mrk421_ASCA_1998}.
Otherwise, the PSD shape of \src\ is possible to be variable in time, 
because the \maxi\ observation does not overlap with the 58-month BAT one.

\subsection{Physical Implications} 
\label{sec:interpretation}
By making most of \maxi\ and \asca,
we confirmed that the PSD of the X-ray variability pattern 
from the VHE blazar \src\ is significantly flatter in the lower frequencies
($f < 2 \times 10^{-6}$ Hz) than 
in the higher frequencies ($f > f_{\rm b}$).
A steep PSD with an index of $\alpha_{\rm H} = 2$ -- $3$ 
in the $f \gtrsim 10 ^{-5}$ Hz range was commonly observed from the VHE blazars 
\citep{Blazar_PSD_ASCA}. 
This index is significantly higher than 
those of the the typical Seyfert galaxies, $\alpha = 1$ -- $2$,
when their PSDs at $f = 10^{-8}$ -- $10^{-4}$ Hz were modeled 
by a single PL function \citep[e.g.,][]{PSD_Simulation_2,Sy_PSD}. 
The steep PSD of the VHE blazars has been widely 
ascribed to the physics operated in their jets,
such as timescales related to the internal shocks
\citep[e.g.,][]{Blazar_PSD_ASCA}.

Interestingly, the \maxi\ result on the PSD slope of \src\ 
below the break frequency ($\alpha_{\rm 0} = 1.60 \pm 0.25$) 
agrees well with those of the typical Seyferts ($\alpha = 1$ -- $2$). 
Seyfert X-ray emission is believed to originate 
in the innermost region 
around the supermassive black hole and its accretion disk.
The jet blobs observed in blazars may ultimately be ejected 
from the innermost disk, triggered by some disk instabilities. 
We thus speculate that the similarity of the PSD index 
over timescales longer than a few days (corresponding to $ f<f_{\rm b}$) 
between blazars and Seyferts implies a physical connection 
between the activities in the jet and the innermost accretion disk
(i.e., the activity of the accretion disk is possible 
to be propagated to the jets). 
However, PSD indices of $\alpha = 1$ -- $2$ are frequently observed 
in various classes of astrophysical objects, 
such as neutron star binaries \citep{VelaX1_PSD}, 
magnetars \citep{magnetar_PSD}, and so forth. 
Obviously, we cannot prove the disk-jet connection
from the PSD shape alone. 
To confirm this scenario,
we must construct a realistic model that simultaneously reproduce
the PSDs of both blazars and Seyferts over a wide frequency range.
Such an undertaking is beyond the scope of the present paper.

\acknowledgments 
The constructive comments from the anonymous referee 
have significantly improved the quality of the present paper.
Thanks to significant support from Dr. Emmanoulopoulos,
the lightcurve simulations satisfying the observed PSD and PDF 
have been made possible.
We are grateful to Dr. M. Kino and Dr. S. Koyama for their fruitful advice.
This research has made use 
of the \maxi\ data \footnote{http://maxi.riken.jp/top/index.php}, 
provided by RIKEN, JAXA and the \maxi\ team.
We acknowledge 
the support from the Ministry of Education, Culture, Sports, 
Science and Technology (MEXT) of Japan, 
through the Grant-in-Aids 22740120 (NI) and  23540265 (YU).



\clearpage 
\begin{table}[h]
\caption{X-ray sources detected within the \src\ region}
\label{table:src}
\begin{center}
\begin{tabular}{lllll}
\hline \hline 
Source & (Ra,      Dec    )      & $\Delta\theta$ \tablenotemark{a} 
                                 & $f_{\rm 4-10}$ \tablenotemark{b}  
                                 & $\sigma$     \tablenotemark{c}  \\
\hline         
\src   & ($166.114$, $+38.209$)  &  ---           
                                 & $1.8 \times 10^{-10}$         
                                 & $107.1$ \\
Src 1  & ($170.282$, $+42.214$)  & $5^\circ.1$    
                                 & $8.8 \times 10^{-12}$         
                                 & $6.0$  \\
Src 2  & ($172.249$, $+36.339$)  & $5^\circ.2$    
                                 & $8.2 \times 10^{-12}$         
                                 & $6.2$  \\ 
\hline         
\end{tabular}
\end{center}
\tablenotetext{a}{Angular separation from \src}
\tablenotetext{b}{The \maxi\ GSC flux in the 4 -- 10 keV range 
expressed in \flux, averaged over the 3 years \citep{2nd_MAXI_catalog}}
\tablenotetext{c}{Source significance}
\end{table}

\clearpage 
\begin{figure*}[h]
\plotone{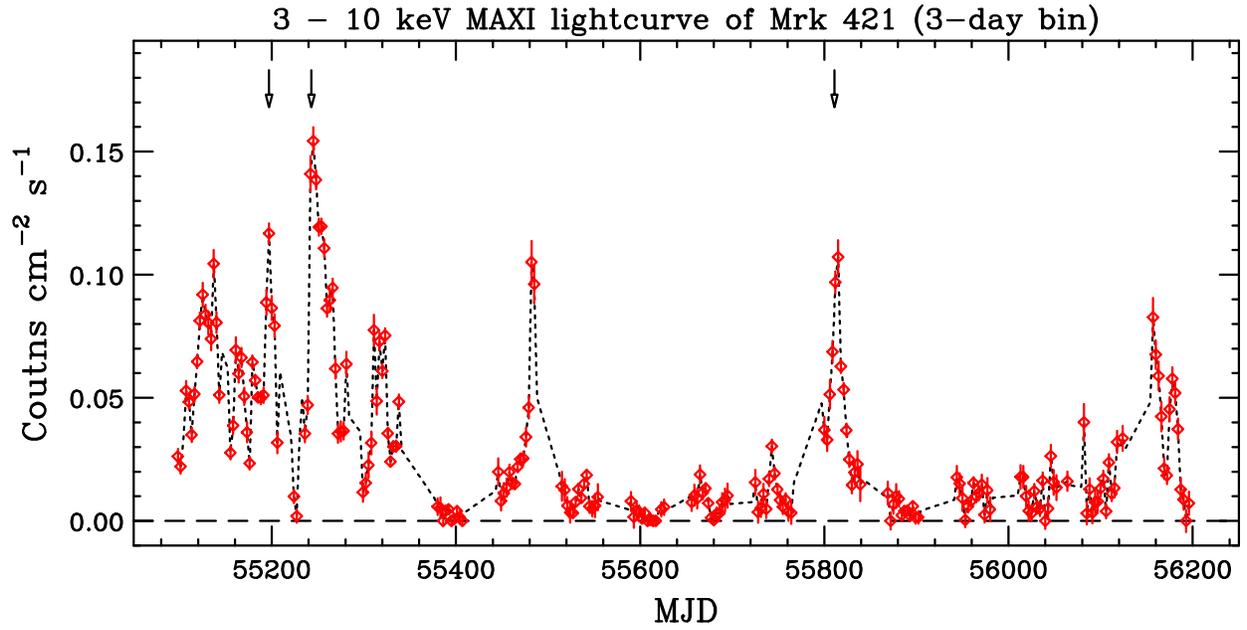}
\caption{
3 -- 10 keV \maxi\ GSC lightcurve of \src, covering 3 years 
between 2009 September 23 (MJD 55097) and 2012 October 15 (MJD 56215).
Time resolution is 3 days.
Red diamonds are data observed from the source.
Unobservable periods (data gaps) have been filled 
by interpolating the real data (dotted line).
The X-ray flares successfully alerted by \maxi\ 
\citep{Mrk421_MAXI,Mrk421_flare_MAXI_3} are indicated by arrows. }
\label{fig:lightcurve}
\end{figure*}

\begin{figure*}[h]
\plotone{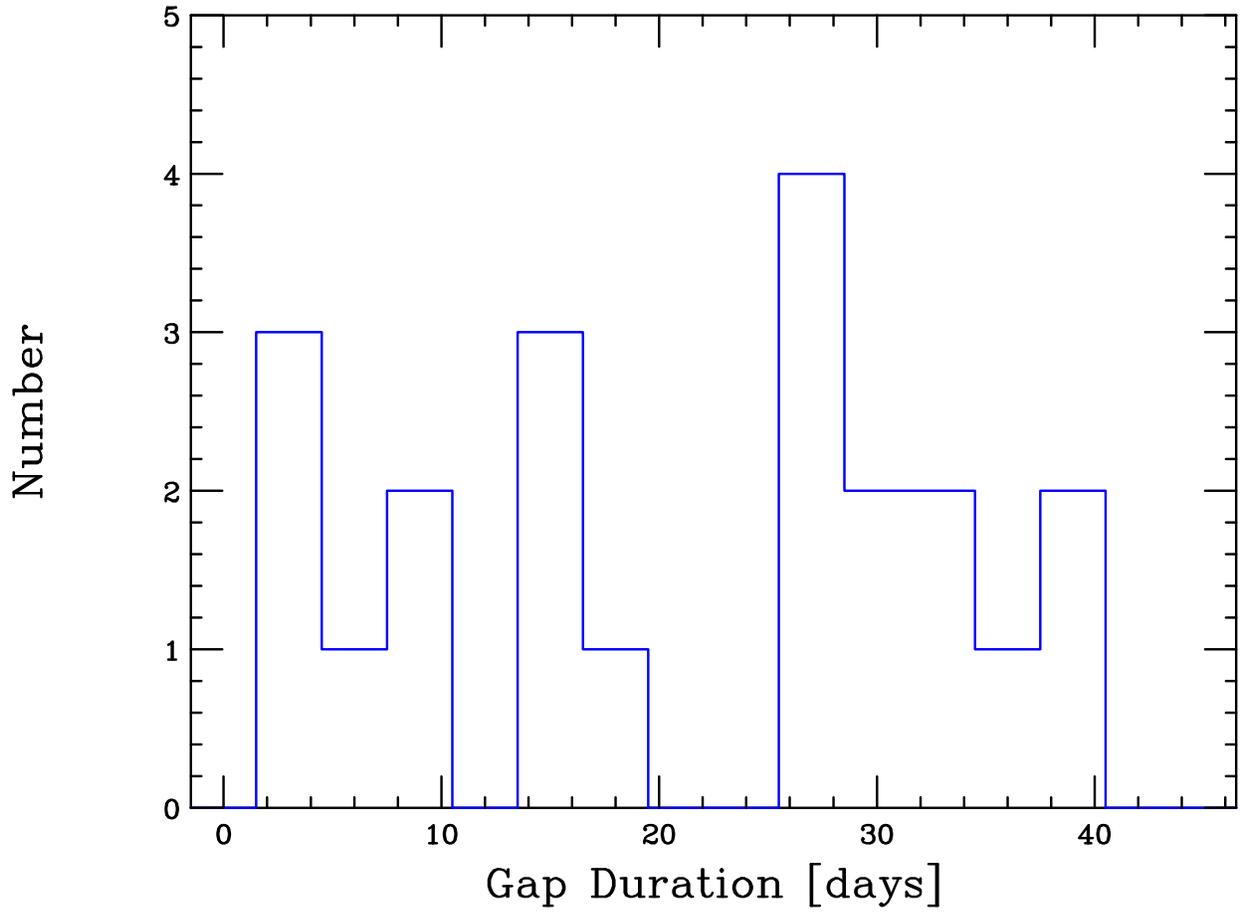}
\caption{Histogram of the data gap duration, 
derived from the 21 data gaps found in the \maxi\ lightcurve 
(Figure \ref{fig:lightcurve}). }
\label{fig:gap_hist}
\end{figure*}

\begin{figure*}[h]
\plottwo{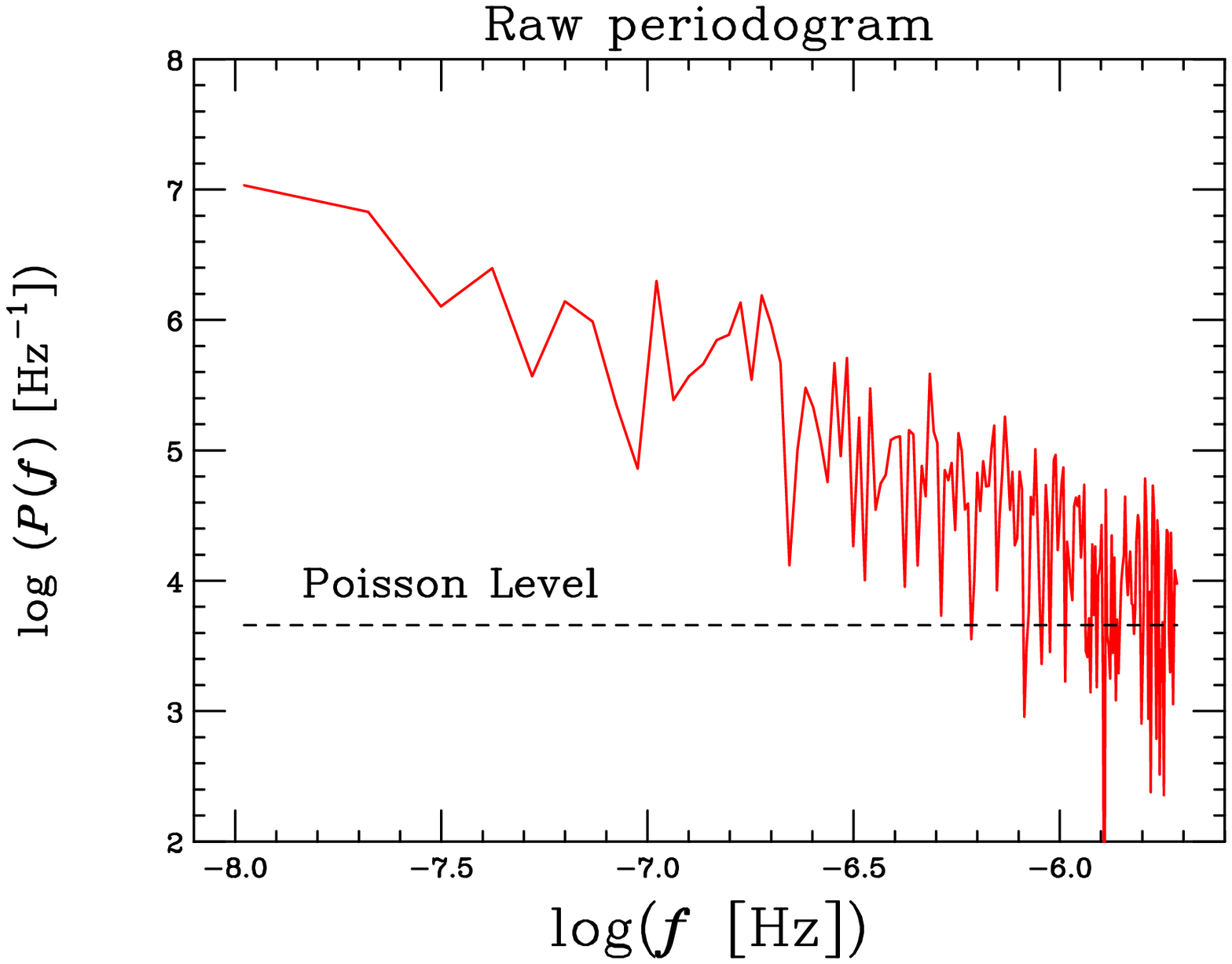}{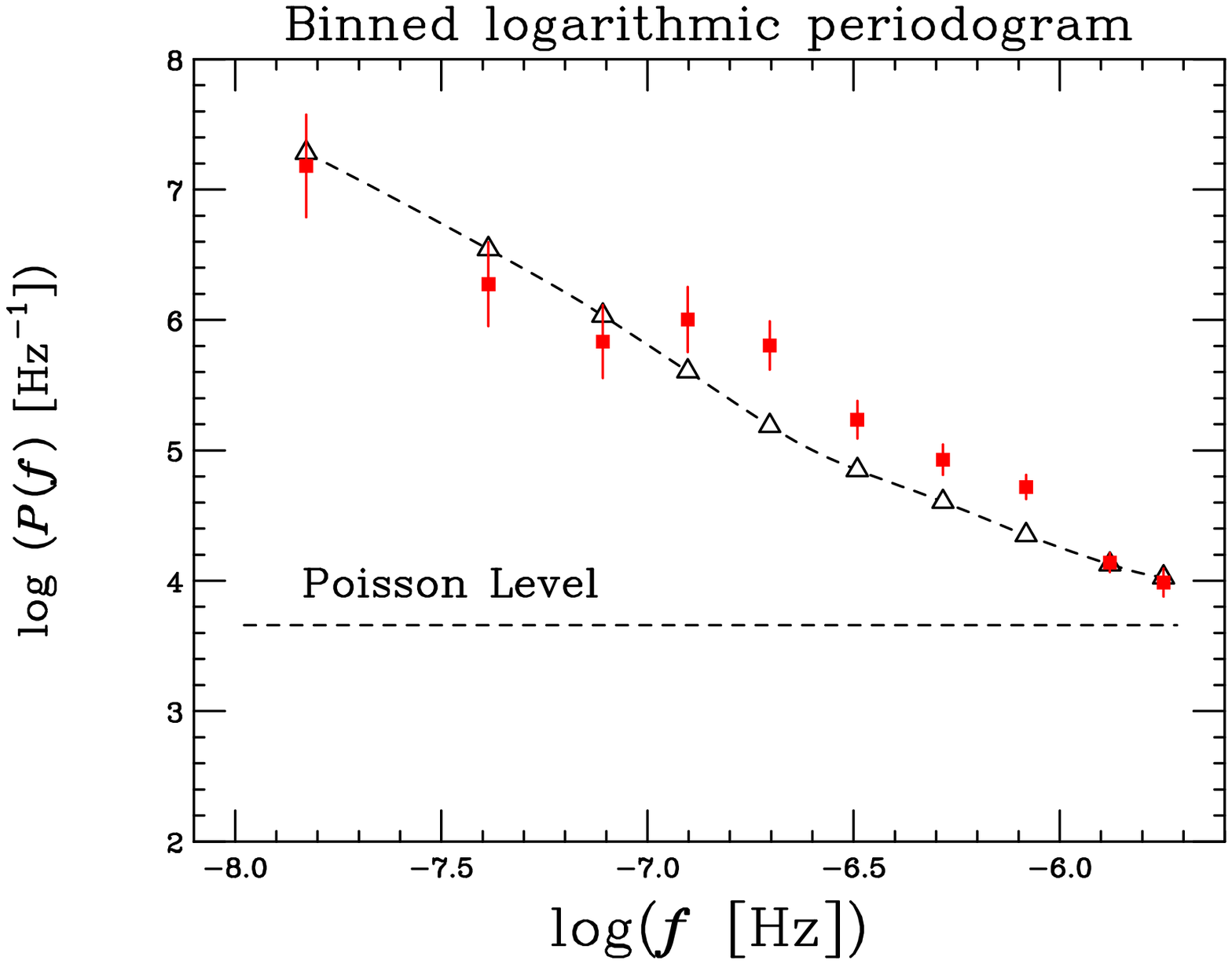}
\caption{
(Left)
Raw periodogram derived from the \maxi\ GSC lightcurve 
of \src\ in Figure \ref{fig:lightcurve}.
The level of the Poisson fluctuation
($P_{\rm stat} = 4.56 \times 10^{3}$ Hz$^{-1}$)
is indicated by the horizontal dashed line.
(Right)
The binned logarithmic periodogram ($\overline{\log(P(f))}$),
evaluated after \citet{red_noise},  
is plotted against the geometric mean frequency for each bin. 
The constant bias to the binned logarithmic periodogram
was corrected by adding 0.25068 to $\overline{\log(P(f))}$ \citep[][]{Vaughan}. 
The periodogram simulation for the best-fit PSD model ($\alpha_{\rm 0} = 1.60$)
is indicated by the open triangles, connected by the dashed line.
Although the Poisson level was unsubtracted from the observed data,
it is taken into account in the simulation.
}
\label{fig:psd}
\end{figure*}

\clearpage 
\appendix
\section{Lightcurve simulation for an arbitrary PSD and PDF} 
The TK95 method, 
which we adopted for the lightcurve simulation,
assumes that the count rate is strictly distributed in a Gaussian function. 
However, as shown in Figure \ref{fig:lightcurve},
\src\ exhibited a bursting behavior,
indicating that the probability density function (PDF) of its lightcurve
significantly deviates from Gaussianity. 
Recently, \citet[][E13]{ArtificialLC} proposed 
a sophisticated technique to generate lightcurves
satisfying simultaneously an arbitrary PSD and PDF.
By comparing the results from these two methods, 
we briefly investigate systematic impacts on the PSD estimate
due to the adopted simulation procedure. 

At this point,
we are only interested in approximately checking
the sensitivity of our results to the lightcurve burstiness.
Thus, we use some simplified forms of the input PSD and PDF, 
just for testing purposes.

Referring to \S \ref{sec:PSDindex}, 
we simply adopted the PL-type underlying PSD model 
described as $S(f) = N_0 (f/f_{\rm b})^{-\alpha_0}$.
In order to generate lightcurves 
with the same statistics as for the observed one, 
the following condition was imposed (see Appendix of E13);
\begin{equation}
\label{eq:PSD-PDF_consistency}
\frac {\sigma^2} {\mu^2} = 2 \int_{f_1}^{f_2} S(f) df, 
\end{equation}
where $\mu = 2.9 \times 10^{-2}$ cts s$^{-1}$ 
and $\sigma = 3.2 \times 10^{-2}$ cts s$^{-1}$, respectively, 
is the unweighted average and standard deviation of the lightcurve
(see \S \ref{sec:photometry}),
$f_1 = 1.05 \times 10^{-8}$ Hz is the lowest frequency of the periodogram
determined from the length of the lightcurve,
and 
$f_2 = 1.93 \times 10^{-6}$ Hz corresponds to the Nyquist frequency 
for the sampling time of $3$ days.  
\footnote{
It is important to note that there is a factor of $2$ difference
in the PSD definitions between the present paper and E13.
Thus, Equation (\ref{eq:PSD-PDF_consistency}) 
contains the factor 2 in front of the integral.
}
Here, we assumed an idealized situation without any Poisson level.

We perform the lightcurve simulation for 3 cases, listed in 
Table \ref{table:artificial_simulation}.
Case 2 represents the best-fit PSD model 
($\alpha_0 = 1.60$; see Figure \ref{fig:prob_reject}),
derived from the 6 data points in the binned logarithmic periodogram,
although its PSD normalization ($N_0 = 6.6 \times 10^{2}$ Hz$^{-1}$)
is different from the best-fit value ($N_0 = 4.4 \times 10^{2}$ Hz$^{-1}$).
There are two reasons for this discrepancy.
The major one is that the hump feature in the observed \maxi\ periodogram
was not considered in the best-fit PSD model
as is visualized in Figure \ref{fig:comp_maxi-asca},
while its power is taken into account in the PSD normalization employed here 
through equation (\ref{eq:PSD-PDF_consistency}). 
The other minor reason is that we neglected the Poisson level,
which also contributes to the observed standard deviation.
The index of Cases 1 and 3 ($\alpha_0 = 1.40$ and $1.80$)
is located near the edge of the acceptable range.
Note that this section contains 
a rough implementation of the E13 method just for testing purposes.

\begin{figure}[h]
\plotone{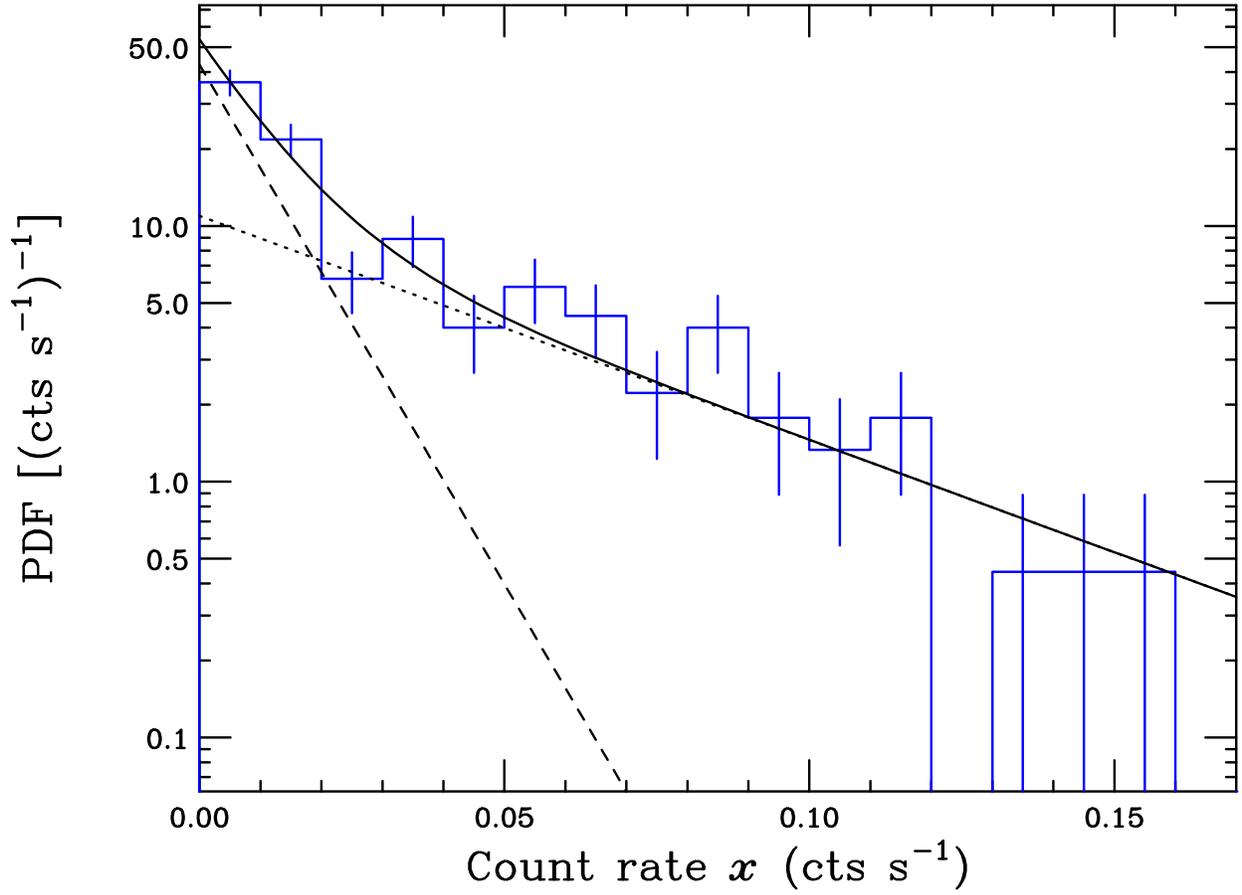}
\caption{PDF of the observed MAXI lightcurve 
         plotted as a function of the count rate $x$.
         The solid line indicates the best-fit model function $p(x)$,
         defined in equation (\ref{eq:PDF_model}).
         The two exponential components of $p(x)$ are
         shown the dotted and dashed lines.}
\label{fig:pdf}
\end{figure}

Figure \ref{fig:pdf} plots the distribution of the count rate $x$,
extracted from the observed \maxi\ lightcurve of \src.
The figure confirms that 
the PDF is significantly far from the Gaussian function.
We approximated the count rate histogram with a PDF model 
described as 
\begin{equation}
\label{eq:PDF_model}
p(x) = A_1 \exp \left( - \frac{x}{B_1} \right) 
     + A_2 \exp \left( - \frac{x}{B_2} \right). 
\end{equation}
The parameters in the function were tied as $A_2 = \frac{1-A_1 B_1}{B_2}$
so that the function meets the condition of $\int p(x) dx = 1$
(i.e., the definition of the PDF).
As is shown with the solid line in Figure \ref{fig:pdf},
the model was found to reasonably reproduce the observed histogram
($\chi^2/{\rm dof} = 1.66$)
with the parameters tabulated in Table \ref{table:pdf}. 
We employed this solution as the input PDF model to 
the lightcurve simulation by the E13 procedure. 
We have to be careful about the fact that 
this PDF model is affected by the data gaps 
(while the gap effect was subtracted from the input PSD model 
as shown in \S \ref{sec:PSDindex}).
It is important to note that 
the standard deviation corresponding to the PDF model 
($\sigma = 4.2 \times 10^{-2}$ cts s$^{-1}$) 
is higher than that of the observed lightcurve,
when we integrate it over the range of $x = 0 $ -- $\infty$.
This is due to the long tail of the PDF function toward higher count rates,
even though its integrated probability at $x > 0.16$ is less than 2\%.
If we limit the count rate at $x \le 0.16$ cts s$^{-1}$,
where the observed data points are distributed, 
the standard deviation becomes consistent to the observed value. 
 
For each parameter set (i.e., Cases 1 -- 3), 
we generated 1000 lightcurves by the E13 technique.
We confirmed that the average count rate and standard deviation 
input to equation (\ref{eq:PSD-PDF_consistency}) 
were reproduced in the simulated products for all the cases.
After the the data gaps in the observed lightcurve are applied,
the individual simulated lightcurves were converted to 
the binned logarithmic periodogram
with the procedure explained in \S \ref{sec:sim_data_gap}.
The ensemble average and standard deviation 
were evaluated from the 1000 binned logarithmic periodograms,
in the similar manner to the TK95 simulation. 

The binned logarithmic periodograms from the E13 and TK95 simulations
are found to coincide with each other within their scatters 
(i.e., the standard deviations).
After the averaged values of 
the binned logarithmic periodogram from these two methods 
were re-converted to linear values,
we found that the E13 result is smaller 
by only a factor of $1$ -- $2$ than the TK95 result for all the cases.  
Therefore, we have concluded that 
our result is relatively insensitive to the adopted simulation technique.

\clearpage

\clearpage
\begin{table}[h]
\caption{Parameters of the underlying PSD model.}
\label{table:artificial_simulation}
\begin{center}
\begin{tabular}{lll}
\hline\hline 
Case &  $\alpha_0$ &  $N_0$ (Hz$^{-1}$)\\ 
\hline 
1    & $1.40$      & $1.9 \times 10^{3} $ \\
2    & $1.60$ \tablenotemark{a} 
                   & $6.6 \times 10^{2} $ \\
3    & $1.80$      & $2.2 \times 10^{2} $ \\
\hline 
\end{tabular}
\tablenotetext{a}{The best-fit PSD index to the observed \maxi\ periodogram}
\end{center}
\end{table}

\begin{table}[h]
\caption{Best-fit PDF parameters}
\label{table:pdf}
\begin{center}
\begin{tabular}{lllll}
\hline\hline 
Parameter & $A_1 $ (cts s$-^{-1}$)$^{-1}$ & $B_1$ cts s$^{-1}$  
          & $A_2 $ (cts s$-^{-1}$)$^{-1}$ & $B_2$ cts s$-^{-1}$ \\
\hline 
Value     & $42.78$                     & $0.0107$ 
          & $10.95$ \tablenotemark{a}   & $0.0495$\\ 
\hline 
\end{tabular} 
\end{center}
\tablenotetext{a}{This parameter was set at $A_2 = \frac{1-A_1 B_1}{B_2}$.}
\end{table}

\end{document}